\documentclass[superscriptaddress,amsmath,amssymb,aps,floatfix,twocolumn,physrev,showkeys,longbibliography]{revtex4-2}
\usepackage{graphicx,dcolumn,bm}
\usepackage{subcaption,amsmath,gensymb,esvect,xcolor,xr,multirow,commath}
\usepackage[colorlinks=true,citecolor=blue]{hyperref}
\usepackage[normalem]{ulem}

\newcommand{\etal}{\textit{et al.}}
\newcommand{\mycomment}[1]{}

\usepackage{etoolbox} 
\usepackage{lipsum} 
\usepackage[capitalize]{cleveref}

\makeatletter
\appto{\appendix}{%
  \@ifstar{\def\theequation@prefix{A.}}%
          {}%
}
\makeatother

\begin{document}

\title{Half-quantized anomalous Hall conductance in topological insulator/ferromagnet van der Waals heterostructures}

\author{Shahid Sattar}
\affiliation{Department of Mathematics and Physics, Linnaeus University, SE-39231 Kalmar, Sweden}

\author{Roman S. Stepanov}
\affiliation{Department of Mathematics and Physics, Linnaeus University, SE-39231 Kalmar, Sweden}

\author{Alexander Tyner}
\affiliation{Nordita, KTH Royal Institute of Technology and Stockholm University,
Hannes Alfv\'{e}ns v\"{a}g 12, 106 91 Stockholm, Sweden} 
\affiliation{Department of Physics, University of Connecticut, Storrs, Connecticut 06269, USA}

\author{M. F. Islam}
\affiliation{Department of Physics, University of Texas at El Paso, El Paso, Texas 79968, USA}

\author{A. H. MacDonald}
\affiliation{Department of Physics, University of Texas at Austin, Austin, Texas 78712, USA}

\author{C. M. Canali}
\affiliation{Department of Mathematics and Physics, Linnaeus University, SE-39231 Kalmar, Sweden}

\date{\today}

\begin{abstract}
The half-quantized anomalous Hall conductance (AHC) in topological materials is a condensed matter physics realization of the parity anomaly of (2+1) quantum field theory and an important challenge for both theoretical and experimental research. A possible realization of this phenomenon may be achieved by interfacing a two-dimensional (2D) ferromagnetic (FM) layer with one surface of a thin slab of a topological insulator (TI), which breaks the otherwise conserved time-reversal symmetry, leading to a gap opening in the Dirac-like energy spectrum of the TI surface states. The resulting heterostructure can support chiral currents where only one spin channel contributes to transport, producing a half-quantized Hall conductance ($e^2/2h$). In this work, using first-principles methods together with tight-binding models, we investigate the magnetization-induced gap, the properties of the sidewalls states, and Hall conductance in three different FI/TI van der Waals heterostructures that are relevant for ongoing experiments. We also discuss the factors that can hinder the realization of exact half-quantization in a realistic system and their implication for the quantum anomalous Hall effect and the topological magnetoelectric effect.
\end{abstract}
\maketitle

\section{Introduction}

Three-dimensional (3D) topological insulators (TIs) are bulk magneto-electric materials characterized by a quantized axion field, whose electromagnetic action is described by the topological term $\mathcal{L}_{\theta}=\frac{\theta}{2\pi}\,\frac{e^2}{hc}\,\mathbf{E\cdot B}$, where $\theta$ is the Chern-Simons magnetoelectric coupling, taking the value $\theta =\pi$ (mod $2\pi$) for strong TIs \cite{qi2008topological,essin2009magnetoelectric,vanderbilt2018berry}. When time-reversal symmetry is locally broken on the surface of a TI film by an out-of-plane exchange field, the 2D gapless linearly dispersive topological surface states acquire an energy (mass) gap at the Dirac point, which simultaneously leads to the breaking of parity symmetry \cite{essin2009magnetoelectric}. This phenomenon can be viewed as condensed matter physics realization of the well-known parity anomaly in ($2$-space + $1$-time)-dimension field theory \cite{PhysRevD.29.2366}{\footnote{In ($2+1$)-dimension field theory, a gauge invariant regularization process, necessary when a classical parity-symmetric theory for massless fermions is quantized, leads inevitably to the breaking of parity.}}. A remarkable consequence of the parity anomaly is the appearance of a parity-violating 2D electric current with a half quantized anomalous Hall conductance, 
$\sigma_{xy}^{\text{local}}=\pm \frac{e^2}{2h}$,
which provides a boundary manifestation of the bulk quantized magneto-electric axion field \cite{chu2011surface,varnava2018surfaces,rauch2018geometric}. 
Indeed, the half-quantized anomalous Hall effect ($1/2$-QAHE) on a magnetized TI surface and the predicted quantized topological magneto-electric (TME) response associated with the quantized axion field are two closely connected phenomena and an example of the bulk-boundary correspondence.
Detecting the effect of the parity anomaly in the form of the $1/2$-QAHE is however non-trivial, since according to the Nielsen-Ninomiya doubling theorem \cite{NIELSEN198120}, Dirac cones in 2D always appear in pairs. Their combined contribution results in the restoration of the parity symmetry, which implies an overall integer QAHE, $\sigma_{xy}^{\text{local}}=\pm n \frac{e^2}{2h}$ with $n = 0, 1, 2 \cdots$. This is precisely the case of TI film, when both surfaces are magnetized, leading to either the Chern insulator phase with an integer QAHE or to the axion insulator phase with zero Hall plateaus and large longitudinal resistance. Nevertheless, the fact that in a TI film the two Dirac cones appear on two distinct top and bottom surfaces, offers the possibility of isolating their half-quantized anomalous Hall contribution \cite{hu2026half}.

The first experimental realization of the parity anomaly state in magnetic TI thin films was reported by Mogi \etal\,at temperatures up to 2K \cite{Mogi2022}. In this work, Cr doping on the top surface of a TI thin film breaks time-reversal symmetry and opens up a magnetic exchange gap in the surface Dirac cone, while the bottom surface remains gapless. Subsequently, Jain \etal\,raised the operating temperature to 10 K by coupling a magnetic CrGe$_2$Te$_6$ layer to top surface of a TI thin film through exfoliation and mechanical stacking \cite{Ralph2024}. In both experiments, the Hall conductivity was observed to lie within a narrow window around the half-quantized value, $\sigma_{xy}=(0.48 - 0.52)$ e$^2/\hbar$. Moreover, the presence of a non-zero longitudinal conductivity, $\sigma_{xx}$, in these experiments is attributed to the conductance arising from the metallic bottom surface states. Recently, Hu \etal\,achieved the layer 1/2-QAHE in a magnetic TI heterostructure, albeit at very low temperature of 30 mK, by positioning the Fermi level within the magnetic gap of the top surface and employed it as a probe of the quantized axion field \cite{hu2026half}. In this context, several works discuss the possibility of achieving magnetization-induced exchange gap at a selected TI surface using MnBi$_2$Te$_4$ \cite{Lu2021,Gu2021,honma2023antiferromagnetic}. 

2D ferromagnets (FMs) having large out-of-plane magnetic anisotropy \cite{gong2017discovery,huang2017layer,lado2017origin} have been proposed and widely used to induce sizable exchange gap in TI/FM van der Waals heterostructures \cite{hou2019magnetizing,yao2019record,li2021quantum,llacsahuanga2022gate,gupta2022gate}. 

Motivated by recent experimental advances, especially on the use of ferromagneticc layer in proximity to a TI in Ref. \cite{Ralph2024} , we considered 2D TI in the form of a six-quintuple-layer (6QL) Bi$_2$Se$_3$ thin film interfaced with three different 2D magnets to investigate magnetization-induced gap, layer-dependent Chern contributions and the resulting topological properties. An important open question in $1/2$-QAHE systems concerns the nature of the associated side-wall edge currents. In contrast to the exponentially localized edge states of the integer QAHE \cite{vandenberghe2017imperfect}, recent theory has predicted that the edge-current distribution in half-quantized systems may exhibit a much slower power-law decay away from the boundary \cite{Zou_2022_PhysRevB.105.L201106}. This also provides a strong motivation to examine how the side-wall current in our vdW heterostructures decays away from the boundary. Finally, owing to the presence of non-zero longitudinal conductivity in experimental works \cite{hu2026half,Ralph2024,Mogi2022}, we also analyze whether metallic states originating from gapless bottom surface in our calculations can account for a finite longitudinal transport response.

The paper is organized as follows: in Section II, we provide computational details and methods employed in this study. 
In Section III, we discuss features of a 6QL-Bi$_2$Se$_3$ thin film which is used as a TI, followed by introducing 2D FM layers. We then discuss electronic structure and topological characteristics of each FM/TI vdW heterostructure in the subsequent sections. In Section IV, we provide results on side-wall edge transport and lastly, we summarized our findings in Section IV.

\section{Computational Details}
We performed first-principles calculations using density functional theory (DFT) with projector augmented waves\,\cite{paw1,paw2} as implemented in the Vienna Ab-initio Simulation Package \cite{vasp}. Generalized gradient approximation in the Perdew-Burke-Ernzerhof parametrization was used to describe the exchange-correlation effects, with a plane wave cutoff energy set to 430 eV. The van der Waals interactions have been taken into account using the DFT-D3 method with Becke-Johnson damping function \cite{grimme2011effect}. A gamma-centered $9\times 9\times 1$ and $12\times 12\times 1$ k-mesh was employed for the structural relaxation and self-consistent calculations, respectively. Due to the involvement of heavy elements, spin-orbit effects were always included in the electronic structure calculations. To obtain iterative solution of the Kohn-Sham equations, we achieved an energy convergence of $10^{-6}$ eV and a force convergence of $10^{-3}$ eV/\AA\, in our calculations. To avoid out-of-plane periodic image interactions, we have also used a 15\,\AA\,thick layer of vacuum. To correctly describe strong on-site Coulomb interactions of localized Mn or Cr $d-$electrons, we used Hubbard $U$ correction (GGA+U method) following Dudarev's scheme \cite{dudarev1998electron} and fixed the onsite Coulomb (U) and exchange (J) parameters to the value of 3.9 eV and 0.0 eV for the case of MnBi$_2$Se$_4$, 3.0 eV and 0.9 eV for CrI$_3$, and 2.0 eV and 0.0 eV for Cr$_2$Ge$_2$Te$_6$, respectively. A real-space tight binding Hamiltonian was obtained based on maximally localized Wannier functions (MLWFs) using the Wannier90 package \cite{marzari1997maximally,mostofi2014updated,pizzi2020wannier90}. For this purpose, we used the VASP2WANNIER90 interface and included Mn/Cr$-d$ orbitals and Bi/Se/Te$-p$ orbitals in generating the Wannier functions. The anomalous Hall conductivity (AHC) was computed within the intrinsic Berry-curvature formalism for slab geometry using the Fermi-sea expression implemented in the WannierBerri package \cite{tsirkin2021high} and is given below:
\begin{equation} 
\begin{aligned}
\sigma_{xy}=- \frac{e^2}{\hbar}
\sum_{n}\int_{BZ}\!\frac{d^2k}{(2\pi)^2} f_{n\mathbf k\,}\Omega_{z}^n(k),
\end{aligned} \label{eq-1}
\end{equation}
where $f_{n\mathbf k}$ is Fermi-Dirac distribution function giving band occupation probability and $\Omega_{z}^n(k)$ is the Berry-curvature of $n$-th band. To analyze the layer-resolved topology, we further compute the partial Chern number $C(l)$ for each orbital $l$, following the definition given in Ref. \cite{varnava2018surfaces} and implemented in pcn package \cite{nicodemosvarnava} via the following expression:
\begin{equation}
C(l)= -\frac{4\pi}{N_k} \mathrm{Im}\sum_{\mathbf{k}}\sum_{v,v' \in \mathrm{occ}}\!\left[
\psi_{v\mathbf{k}}(l)\,
\psi^{*}_{v'\mathbf{k}}(l)\,
F_{vv'}(\mathbf{k})
\right],
\end{equation} whereas $\psi_{v\mathbf{k}}(l)\,
\psi^{*}_{v'\mathbf{k}}(l)$ is the wavefunction amplitude at crystal momentum $k$ on each orbital $l$,  $F_{vv'}(\mathbf{k})$ denotes the covariant-curvature matrix in the occupied-band subspace, constructed from virtual transitions between occupied and unoccupied states and occupied bands labeled as $\nu$ and $\nu'$. The layer-resolved Chern number is then obtained by summing $C(l)$ over all orbitals belonging to the corresponding given layer.

\begin{figure*}
\centering
\includegraphics[width=0.95\linewidth]{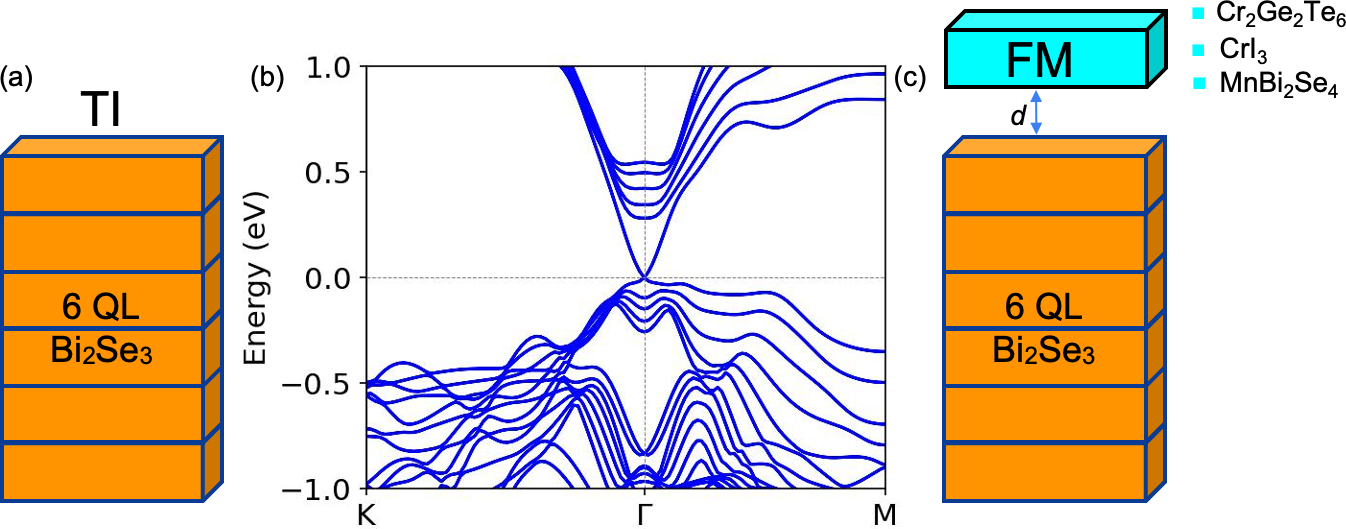}
\caption{\label{fig1} (a) Schematic of a pristine six-quintuple-layer (6QL) Bi$_2$Se$_3$ thin film used as a topological insulator (TI). (b) Calculated electronic dispersion along the high-symmetry $K\rightarrow \Gamma \rightarrow M$ path in Brillouin zone, showing the gapless Dirac cone near the Fermi level at $\Gamma$-point. (c) Schematic of TI/FM vdW heterostructure geometry considered in this work, where a ferromagnetic layer is placed at a distance $d$ from the TI surface. The three ferromagnetic materials considered are monolayer Cr$_2$Ge$_2$Te$_6$ or CrI$_3$ or MnBi$_2$Te$_4$.}
\end{figure*}

\section{Results and Discussion}

\subsection{Six Quintuple Layer Bi$_2$Se$_3$ thin film as a TI and ferromagnetic candidates}

Figure \ref{fig1}(a) shows the schematic of a pristine six-quintuple-layer (6QL) Bi$_2$Se$_3$ used as a topological insulator (TI) building block. Here, each rectangular block represents one QL stacked along the out-of-plane direction. The calculated band structure of the pristine 6QL slab, shown in panel (b), clearly demonstrates the existence of Dirac-like surface states at the high-symmetry $\Gamma$-point, a hallmark feature of 2D topological insulators. Importantly, DFT-D3 vdW correction with Becke-Johnson damping function \cite{grimme2011effect} provides a band gap of $\sim 1.1$ meV, which is smaller than what is typically obtained using other van der Waals functionals \cite{reid2020first}. It is pertinent to mention that obtaining a small bandgap for slab geometry is crucial for isolating the topological surface states and enabling robust proximity effects in heterostructures. Figure \ref{fig1}(c) depicts the heterostructure setup adopted in the present study, where a FM layer is placed in proximity to the 6QL Bi$_2$Se$_3$ slab at an interlayer distance $d$ specific to each FM layer. Three different 2D FM candidates are considered, i.e., monolayer Cr$_2$Ge$_2$Te$_6$ (CGT), CrI$_3$ (CrI), and MnBi$_2$Se$_4$ (MBSe). This setup allows us to examine time-reversal symmetry breaking at the TI top surface via magnetic-exchange and inducing a finite gap in one of the two Dirac cones coming from top and bottom surfaces.

\begin{figure*}
\centering
\includegraphics[width=0.95\linewidth]{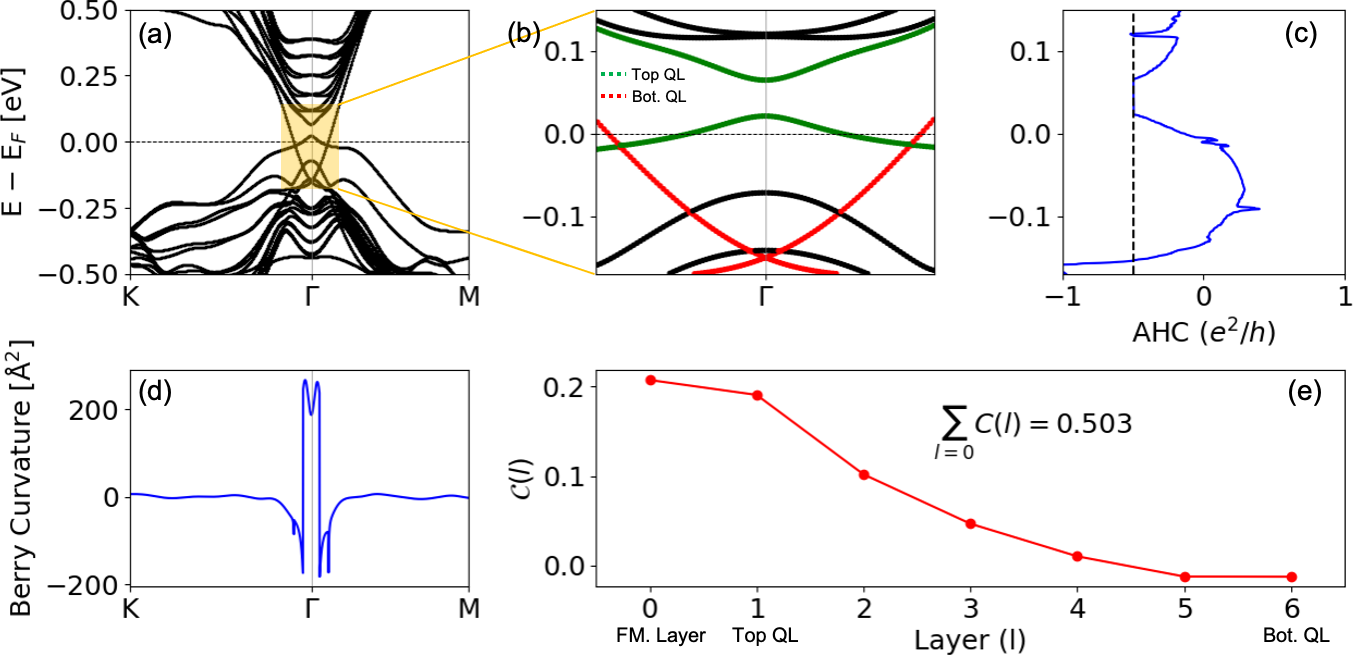}
\caption{(a) Band structure of vdW heterostructure using Cr$_2$Ge$_2$Te$_6$ (CGT) as a ferromagnetic layer on top of the 6-QL Bi$_2$Se$_3$ thin film. (b) Zoom region showing energy-gap between bands of top QL (in green color) and ungaped bands of bottom QL (in red color), respectively (c) AHC as a function of Fermi energy. (d) Berry curvature distribution showing a sharp peak at the high-symmetry $\Gamma$-point in the Brillouin zone. (e) Layer-projected Chern contributions across the CGT/TI heterostructure.}
\label{fig2}
\end{figure*}

\subsection{Cr$_2$Ge$_2$Te$_6$ as a ferromagnetic top layer}

Analogous to the experimental setup adopted in Ref. \cite{Ralph2024} to realize $1/2$-QAHE, the TI/CGT heterostructure provides a natural starting point for our discussion. A key feature of this heterostructure is the vdW contact between the top chalcogen layer of the TI (Se) and the surface chalcogen layer of CGT (Te). In this setup, the spatially extended and highly polarizable Te-$5p$ valence states, compared with the Se-$4p$ states, promote enhanced interfacial charge redistribution and hybridization in the vdW gap. An interlayer distance ($d=2.91$) supports these observations. 

Figure \ref{fig2} shows key features of the TI/CGT heterostructure. As shown in Figure \ref{fig2}(a), the low-energy bands near the Fermi level are strongly modified around the $\Gamma$-point compared with those of the pristine Bi$_2$Se$_3$ slab. The highlighted region marks the topological surface-state manifold, whose detailed structure is enlarged in panel (b). The layer projections shown here distinguish contributions coming from the top QL (green color) and the bottom QL (red color), respectively. A clear asymmetry is observed between the two surfaces: owing to hybridization with CGT, the top surface states are shifted away from the Dirac crossing and become gapped, whereas the bottom surface bands remain nearly gapless and retain an approximately Dirac-like dispersion. This demonstrates that proximity from CGT acts predominantly on the top surface of TI and breaks time-reversal symmetry locally. We observe a bandgap opening of 45 meV in the top surface which is direct consequence of strong interlayer interactions between CGT and top-most QL of Bi$_2$Se$_3$. 

Figure \ref{fig2}(c) shows AHC computed as a function of Fermi energy for TI/CGT heterostructure. A striking feature is the emergence of Hall plateau, $\sigma_{xy}=\frac{e^2}{2\hbar}$, within the energy range of gapped top surface. Combined with QL-projected band structure in Figure \ref{fig2}(b), this clearly demonstrates that a $1/2$-QAHE originates from the exchange-induced gapped top surface of Bi$_2$Se$_3$. The bottom surface, however, does not contribute to AHC owing to gapless nature of surface states. The metallic nature of bottom surface is also confirmed by computing the longitudinal conductivity ($\sigma_{xx}$ and $\sigma_{yy}$), as shown in Supplementary Fig. S1, having non-zero contributions. Both components exhibit a pronounced dependence with nonmonotonic features in the energy range corresponding to the exchange gap. Overall, this scenario corresponds to a $1/2$-QAHE regime where a quantized transverse response coexists with a finite longitudinal conductivity (without destroying the quantization), allowing one to describe the system as exhibiting half-quantized AHC localized on the top surface.

To gain further insight into the origin of $1/2$-QAHE, we next analyze the Berry curvature (BC), as shown in Figure \ref{fig2}(d), which exhibits pronounced peaks concentrated in a narrow region of $k$-space near the $\Gamma$-point and nearly vanishes away from this region. Firstly, this indicates that the dominant contribution to the AHC arises from the low-energy electronic dispersions near the Fermi energy. Secondly, the BC sharp peaks mainly originate near the band crossings/anticrossings around the $\Gamma$-point because of interband mixing and small energy denominators, consistent with earlier studies \cite{nagaosa2010anomalous}. However, in contrast to previous works, where such BC peaks are attributed to hybridization between states of top and bottom surface \cite{zou2023half}, in the present case they originate primarily due to hybridization of the top surface states with FM layer.

Finally, we examine the sum of orbital-resoled layer-projected Chern contributions $C(l)$, shown in Figure \ref{fig2}(e), which provides spatial distribution of the $1/2$-QAHE. The maximal contribution is coming from the proximitized FM and the top-most QL of Bi$_2$Se$_3$ followed by a gradually decrease toward the bottom of the slab. This clearly indicates that the topological $1/2$-QAHE response is localized near the FM/TI interface. The total summed value ($\sum_l C(l)=0.503$) is in agreement with half-quantized AHC obtained in Figure \ref{fig2}(c).

\begin{figure*}
\centering
\includegraphics[width=0.95\linewidth]{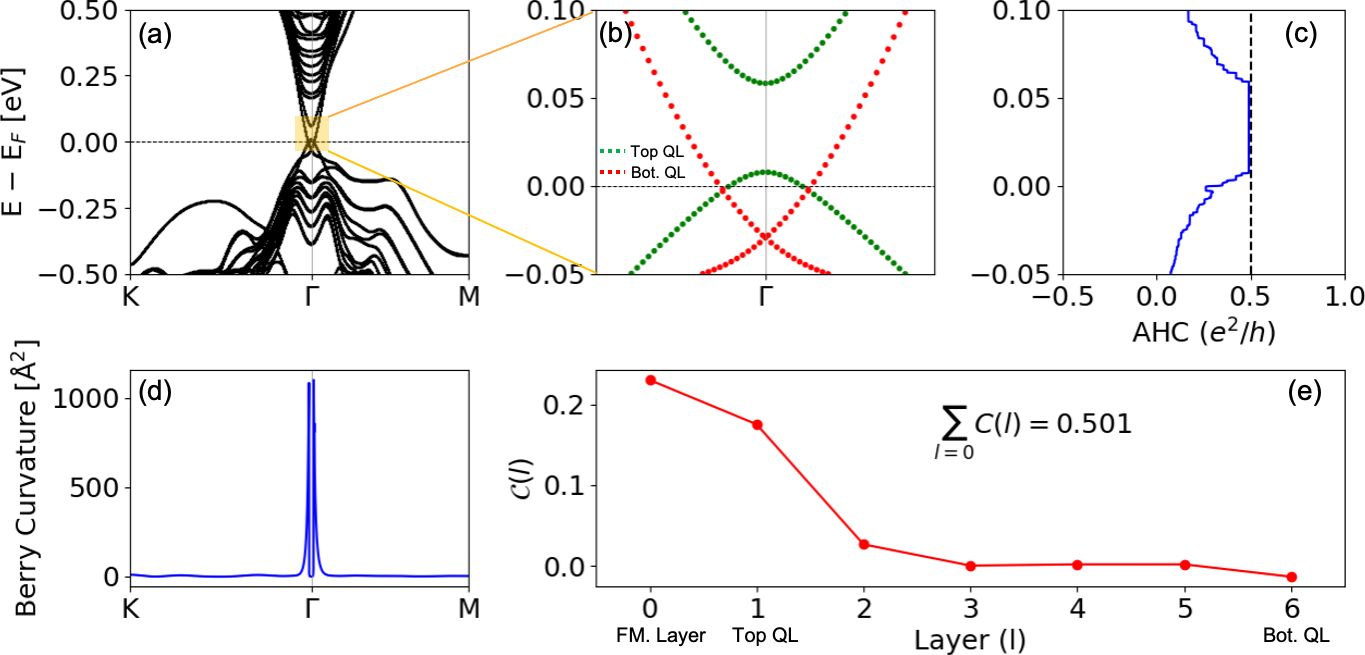}
\caption{(a) Band structure of vdW heterostructure using MnBi$_2$Se$_4$ (MBSe) as a ferromagnetic top layer on top of 6 QL Bi$_2$Se$_3$ thin film. (b) Zoom region showing energy-gap between bands of top QL (in green color) and ungapped bands of bottom QL (in red color). (c) AHC as a function of Fermi energy. (d) Berry curvature distribution along the $K\rightarrow \Gamma \rightarrow M$ high-symmetry path in the Brillouin zone. (e) Layer-projected Chern contributions across the MBSe/TI heterostructure.}
\label{fig3}
\end{figure*}

\subsection{MnBi$_2$Se$_4$ as a ferromagnetic top layer}

We next consider single-layer MnBi$_2$Se$_4$ (MBSe) as a FM contact to the 6 QL Bi$_2$Se$_3$ thin film, even though bulk MBSe is known to exhibit antiferromagnetic order. The resulting MnBi$_2$Se$_4$/Bi$_2$Se$_3$ heterostructure has not yet been realized experimentally, but it can be a promising platform for achieving a half-quantized Hall response. With an interlayer distance of $d=2.67\,$\AA, the interfacial interactions are dominated by Se atoms on both sides of the contact. The resulting charge-density difference here will be primarily determined by a redistribution of Se-$4p$ states. In contrast to the earlier Se–Te contact, this Se–Se charge environment produces a weaker interface step in the electrostatic potential and consequently a less pronounced Fermi-level shift relative to the Dirac states of the bottom TI surface.

Figure~\ref{fig3} presents electronic dispersions corresponding to MBSe/Bi$_2$Se$_3$ vdW heterostructure. Refer to the band structure near the $\Gamma$ point (magnified view of Fig.~\ref{fig3}(a) in panel (b)), two groups of surface bands are clearly resolved, associated with the top QL (green color) and bottom QL (red color) of the TI sample, respectively. Similar to the CGT case, proximity effect by the magnetic layer induces sizable exchange splittings and opens up a gap at the top surface, while the bottom surface remains essentially gapless. Overall, the system is metallic (shown in Fig.~\ref{fig3}(b)). On the other hand, MnBi$_2$Se$_4$/Bi$_2$Se$_3$ vdW heterostructure also experience a half-quantized Hall response given in Figure ~\ref{fig3}(c). Similar to the previous case, the metallic character of electronic dispersion is preserved due to a parallel conducting channel on the bottom surface.

The longitudinal response, given in Supplementary Figure S1(b), confirms the gapless nature of the heterostructure. The longitudinal conductivity ($\sigma_{xx}$) increases when the Fermi energy is moved to positive domain and shows a pronounced non-monotonic behavior. This trend is typical of a regime where additional (interface-related) states become active, leading to changes in the Fermi-surface area, group velocities, and effective anisotropy, and consequently enhancing the longitudinal conductivity.

Despite the presence of a dissipative channel, the AHC ($\sigma_{xy}$), given in Fig.~\ref{fig3}(c), remains close to the half-quantized value ($\approx e^2/2\hbar$), indicating that the response is dominated by the topological contribution from the gaped top surface. Microscopically, this is consistent with Fig.~\ref{fig3}(d), where the Berry curvature ($\Omega_z(\mathbf{k})$) is concentrated in a narrow region of k-space near $\Gamma$, pointing to a localized origin of the main contribution to $\sigma_{xy}$ due to band crossings. Finally, the layer-resolved Chern contributions ($C(l)$) given in Fig.~\ref{fig3}(e) show maximal effect at the top interface and rapidly decays into the subsequent QLs of the TI. This confirms our earlier observation that the $1/2$-QAHE is localized near the top surface, close to the FM overlayer, whereas the gapless bottom surface provides a parallel metallic channel.

\begin{figure*}
\centering
\includegraphics[width=0.95\linewidth]{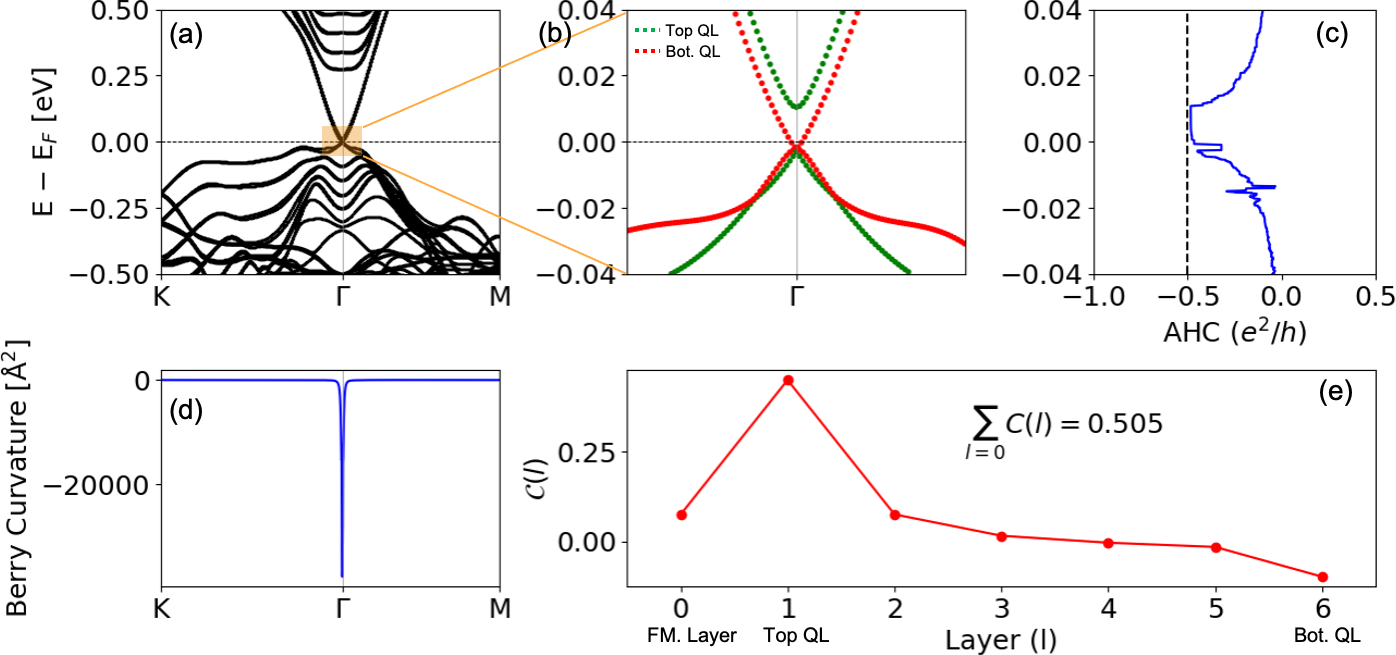}
\caption{Band structure of vdW heterostructure using CrI$_3$ as a ferromagnetic top layer on top of 6 QL Bi$_2$Se$_3$ thin film. (b) Zoom region showing energy-gap between bands of top QL (in green color) and ungapped bands of bottom QL (in red color). (c) AHC as a function of Fermi energy. Berry curvature distribution along the $K\rightarrow \Gamma \rightarrow M$ high-symmetry path in the Brillouin zone. (e) Layer-projected Chern contributions across the CrI$_3$/TI heterostructure.}
\label{fig4}
\end{figure*}

\subsection{CrI$_3$ as a ferromagnetic top layer}

In Fig.~\ref{fig4}(a), we present results for the CrI$_3$/6QL Bi$_2$Se$_3$ heterostructure, similar to the two previous cases. The enlarged band structure near the Fermi energy and around $\Gamma$ point (see Fig.~\ref{fig4}(b)) reveals two groups of surface bands associated with the top QL (in green) and bottom QL (in red) of the heterostructure. Proximity to the magnetic layer induces an exchange splitting in the surface state of the top QL, despite the relatively large interlayer distance $d=3.03\,\AA$, and opens up a gap, whereas the bottom surface states retains a Dirac-like dispersion near the Fermi energy. Thus, as in the previous heterostructures, the system remains metallic due to a parallel conducting channel at the bottom surface. A distinctive feature compared to CGT and MnBiSe is the bottom surface bands forming a small Fermi-surface pocket. Consequently, their contribution to longitudinal transport is suppressed, as shown in Supplementary Information Fig. S2(c), and forms a minima in both $\sigma_{xx}$ and $\sigma_{yy}$, with ($\sigma_{xx}\approx\sigma_{yy}$) indicating weak in-plane anisotropy.

Despite the presence of a metallic channel, the transverse response remains nearly quantized (Fig.~\ref{fig4}(c)) at ($\approx -e^2/2\hbar$). The residual deviation is not caused by a finite ($\sigma_{xx}$), but rather due to additional contributions to ($\sigma_{xy}$) from interface subbands that depend on the details of orbital hybridization. Interestingly, the Berry curvature ($\Omega_z(\mathbf{k})$) in Fig.~\ref{fig4}(d) seems to be concentrated within an extremely narrow region around ($\Gamma$). Unlike in the previous two FM/TI heterostructures, this can be understood from the absence of multiple low-energy band and avoided crossings in CrI$_3$/TI band structure. Finally, the layer-resolved decomposition (Fig.~\ref{fig4}(e)) shows that (C(l)) is maximal near the top interface and rapidly decays into the film thickness, while the total value ($\sum_l C(l)=0.505$) is essentially equal to (1/2). This indicates that the topological contribution is localized at the topmost QL and is not related to bulk of the sample, whereas the bottom surface primarily provides the parallel metallic channel. The negative sign of ($\sigma_{xy}$) is determined by the chosen magnetization orientation (the sign of the exchange term) and the corresponding chirality of the Hall response.

\section{Sidewall states of heterostructure nanoribbon}
\begin{figure*}
\centering
\includegraphics[width=0.95\linewidth]{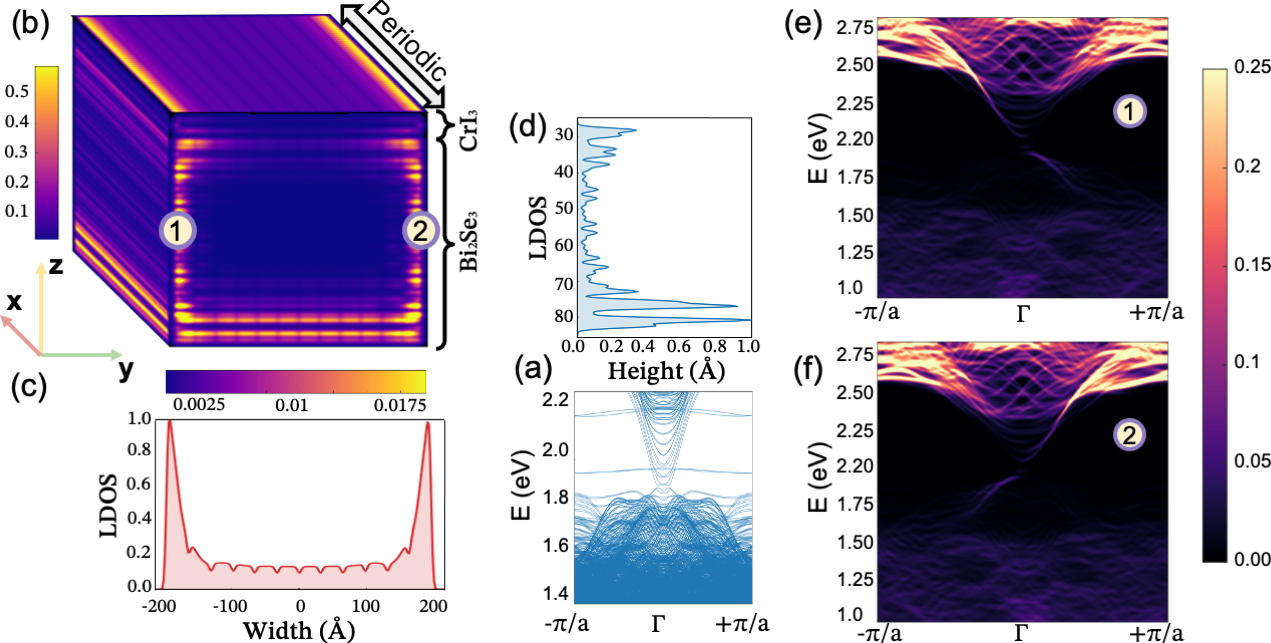}
\caption{(a) band structure of a quasi-1D nanoribbon of a vdW heterostructure consisting of a CrI$_3$ a FM monolayer coupled to a 6 QL Bi$_2$Se$_3$ thin film. (b-d) Spatial distribution of the local density of states (LDOS) for the nanoribbon geometry, presented in three projections: (b) cross-sectional view; (c) top view, as a function of the horizontal (y) position; (d) along the periodic (x) direction, as a function of the vertical (z) position. The arrow indicates the periodic direction along the x-axis, while no periodicity is present along the y- and z-axes. Numbers 1 and 2 denote the position of the localized states at the left and the right edge, respectively. ((e,f) Sidewall-state dispersion for the 1-left and 2-right sidewalls: these band structures are obtained by projecting the full quasi-1D band-structure of the nanoribbon onto the left and right sidewalls respectively. Note the feeble left-moving and right moving 1D states located respectively on the left (1) and right (2) sidewalls.}
\label{fig5} 
\end{figure*}

We conclude our study of the FM/TI heterostructures with the important analysis of the conducting electronic states present in a nanoribbon of finite width carved out of the infinite 2D films considered in the previous sections. Specifically our results are shown in Fig.~\ref{fig5} for a 20 nm-wide naoribbon of the vdW heterostructure consisting of a FM CrI$_3$ monolayer positioned on top of a 6-QL Bi$_2$Se$_3$ TI thin film. In Fig.~\ref{fig5}(a) we plot the quasi 1D band structure for the nanoribbon. The Fermi energy $E_F = 1.82 eV$ is positioned inside the gap of the top (magnetized) Dirac cone and crosses the conducting states of the gapless bottom Dirac cone. Panel (b) shows the local density of states LDOS at the Fermi energy on the cross-section of a nanoribbon. Panel (c) plots instead the top-view LDOS, as a function of nanoribbon horizontal (y) position, while (d) plots the sidewall LDOS (along the periodic direction x), as function of the vertical (z) position. 
Focusing specifically on panel (b) starting from the bottom of the ribbon, we can clearly distinguish the topological quasi-2D states coming from the gapless Dirac cone of the un-magnetized surface. Moving up, non-topological edge states strongly pinned close to the left and right sidewalls of the ribbon. Approaching the magnetized top surface with a gapped Dirac cone, the nature of these sidewall states changes. Here they are characterized by a gradual decay into the inner part of the ribbon that is slower than for the lower SLs; they are spin-polarized and chiral, as it is shown in panels (d) and (e) where we plot the dispersions of ribbon quasi-1D states projected onto the left and right side-walls surfaces. In a Chern insulator phase (of a FM/TI/FM three-layer heterostructure) together with their counterpart coming from the equally magnetized bottom layers, these state generate the chiral states responsible for the integer QAHE. In the present case, although they are not ``proper" topological chiral edge states, they give rise to spin polarized chiral currents which lead to the 1/2 QAHE of the FM/TI heterostructure. It is indeed remarkable that these states, albeit do not possess the same strong topological character of the Chern insulator chiral edge states, in the end lead to the exact half-quantization of the parity anomaly phase.

\section{Conclusion}
In conclusion, by using first-principles methods combined with microscopic tight-binding models we have investigated, the electronic, magnetic and transport poperies of three different van der Waals heterostructures consisting of a topological insulator thin film and a ferromagnetic layer positioned of top of the TI surface. Using first-principles methods together with tight-binding models, we have shown that, for the TI surface where time-reversal symmetry is broken by the contiguous FM layer a gap open up at the Dirac cone of the 2D topological surface state, while the other Dirac cone remains gapless. The size of the exchange gap (of the order a few tens meV) and the relative position of the gapped Dirac cone with respect to gapless cone depend on the FM layer oof the heterostructure. In any case, the Fermi energy of the  heterostructure always crosses the energy band of the gapless Dirac cone, which therefore will always provide dissipative conducting sates to the overall transport properties of the system. 

A calculation of the AHC for an infinite 2D heterostructures (no sidewall) shows that AHC is indeed exactly half-quantized when the Fermi energy is positioned inside the exchange gap of the Dirac states of the magnetized surface, regardless of the presence of the dissipative conducting sates coming from the gapless surface. However, we argued that, although these states do not contribute to the Hall response, they will indirectly affect the exact quantization since they do contribute to the longitudinal resistance, which remains finite. These theoretical results are, in fact, consistent with and explain the experimental findings \cite{Mogi2022,Ralph2024,hu2026half} that the AHC in our studied FM/TI vdW heterostructure is half quantized within a few-percent accuracy.

Our theoretical analysis of the surface states of a quasi-1D heterostructure nanoribbon further demonstrates the presence of the conducting surfaces states of the un-magnetized TI surface. Importantly, this analysis also elucidates the crucial role played by the {\it sidewall states close to the magnetized surface} in establishing spin-polarized chiral currents slowly decaying inside the bulk of the nanoribbon, which are ultimately the physical origin for the half quantization of the Hall conductance characterizing the parity anomaly phase.

\bibliography{main}

\end{document}